\documentstyle[12pt]{article}

\def\o{\over}
\def\b{\begin{equation}}
\def\e{\end{equation}}
\def\l{\label}
\def\kpnn{K^+\rightarrow\pi^+\nu\bar\nu }

\def\klpnn{K_L\rightarrow\pi^0\nu\bar\nu}

\begin{document}
\thispagestyle{empty}
\begin{flushright}
 MPI-PhT/94-25 \\
 TUM-T31-61/94 \\
 May 1994
\end{flushright}
\vskip1truecm
\centerline{\Large\bf  Precise Determinations of the CKM Matrix}
\centerline{\Large\bf From CP-Asymmetries in B-Decays and $\klpnn$
   \footnote[1]{\noindent
   Supported by the German
   Bundesministerium f\"ur Forschung und Technologie under contract
   06 TM 732 and by the CEC science project SC1--CT91--0729.}}
\vskip1truecm
\centerline{\sc Andrzej J. Buras}
\bigskip
\centerline{\sl Technische Universit\"at M\"unchen, Physik Department}
\centerline{\sl D-85748 Garching, Germany}
\vskip0.6truecm
\centerline{\sl Max-Planck-Institut f\"ur Physik}
\centerline{\sl  -- Werner-Heisenberg-Institut --}
\centerline{\sl F\"ohringer Ring 6, D-80805 M\"unchen, Germany}

\vskip1truecm
\centerline{\bf Abstract}
We point out that the measurements of the CP asymmetries in neutral
B-decays together with a measurement of $Br(K_L\to \pi^\circ\nu\bar\nu)$
and the known value of $\mid V_{us}\mid $ can determine {\it all} elements
of the Cabibbo-Kobayashi-Maskawa matrix without essentially any hadronic
 uncertainties.
In particular the strong dependence of $Br(K_L\to \pi^\circ\nu\bar\nu)$
on $\mid V_{cb}\mid$ ($\mid V_{cb}\mid^4$) can be used to determine this
element precisely. Consequently $\mid V_{ub}\mid$, $\mid V_{td}\mid$
and $\mid V_{ts} \mid $ can also be determined.  We derive a set
of formulae which facilitate this study and we present a numerical analysis
of several future scenarios. Accurate
determinations of all parameters are expected from the B-factories and the
Main Injector era at Fermilab including KAMI.
An impressive accuracy should be reached in the LHC era.
An analysis using $Br(\kpnn)$ instead of $Br(\klpnn)$ shows that although
the determination of $\mid V_{cb}\mid$ and
$\mid V_{td}\mid$ this way is less accurate, still useful results for
these elements can be obtained.
\vfill
\newpage

\pagenumbering{arabic}

The Cabibbo-Kobayashi-Maskawa matrix \cite{CAB,KM} which parametrizes
 the weak charged
current interactions of quarks contains four parameters which must be
determined by comparing the theoretical branching ratios
with the experimental data. One of these parameters, the element $V_{us}$,
is known from $K^+\to\pi^\circ e^+\nu$, $K_L\to \pi^- e^+\nu$ and
semileptonic hyperon decays with a high precision:
 $\mid V_{us}\mid\equiv \lambda=0.2205\pm 0.0018$ \cite{LR,DHK}.
Recent critical discussions of this determination and of the related element
$\mid V_{ud}\mid$ can be found in \cite{MAR}.
 The present determinations of the remaining
 three parameters are
subject to theoretical uncertainties resulting from our inability to
perform precise non-perturbative calculations of various hadronic matrix
elements of weak currents and local four-quark operators which enter the
relevant formulae.

It has been stressed in numerous papers
\cite{NQ,AKL} that
CP asymmetries in $B^0\to f$, where $f$ is a CP eigenstate can determine
two parameters $\varrho$ and $\eta$ in the Wolfenstein parametrization
\cite{WO}
without essentially any hadronic uncertainties. A recent analysis of the
related unitarity triangle can be found in \cite{BLO}, where a high
accuracy of this determination at future B-factories has been emphasized.
 Here we would like to
point out that the measurement of the purely short distance CP violating
decay
$K_L\to \pi^\circ\nu\bar\nu$ \cite{LI} together with the CP asymmetries in
question, allows a clean and precise determination of the fourth parameter
$\mid V_{cb} \mid$ (or $A$). The argument is as follows.

The last year calculations
\cite{BB1,BB2} of next-to-leading QCD corrections to \\  $Br(\klpnn)$
 reduced the theoretical uncertainty
due to the choice of the renormalization scales present in the
leading order expressions \cite{DDG} below $\pm 1\%$. Because
 the relevant hadronic matrix
element of the weak current $\bar {s} \gamma_{\mu} (1- \gamma _{5})d $ can
be measured to better than $1 \%$ in the leading
decay $K^+ \rightarrow \pi^0 e^+ \nu$, the resulting theoretical
expression for  $Br(\klpnn)$ is
only a function of $\eta$,
 $V_{cb}$ and $m_t=\overline{m_t}(m_t)$,
the running top quark mass at the $m_t$ scale. The very weak dependence on the
QCD scale  $\Lambda \overline{_{MS}}$ can be safely neglected.
Now as shown below $\eta$  will be known with high precision from future
B-factories \cite{BAB}, HERA-B \cite{AL}, Tevatron with the Main Injector
\cite{FNAL} and LHC
 B-physics experiments \cite{CA,B93}. Similarly Tevatron
 and LHC will offer precise determinations
of $m_t$. Consequently $\mid V_{cb}\mid$ can be determined
 by measuring $Br(\klpnn)$.
Since in addition $Br(\klpnn)$ is a very sensitive function
of $\mid V_{cb}\mid$ ($\mid V_{cb}\mid^4$), the latter element of the
CKM matrix can be extracted from $\klpnn$ with a precision comparable
or even better than achieved directly in tree level semi-leptonic
B-decays.
To this end $Br(\klpnn)$ must be known to within $5 \%$.
Since the present standard model estimate of $Br(\klpnn)$ is in the
ball park of $(20-40)\cdot 10^{-12}$, the sensitivity of
${\cal O}(10^{-12})$ aimed for at KAMI \cite{AR} and KEK \cite{ISS}
should allow to
achieve this goal.

At first sight it is probably surprising that we use a rare K-meson
decay to determine $\mid V_{cb}\mid$. The natural place to do this
are of course tree level B-decays. On the other hand using unitarity
and the Wolfenstein parametrization with $\mid V_{cb}\mid=A\lambda^2$
it is clear that $\mid V_{cb}\mid$ gives the overall scale $A$ of the
top quark couplings $V_{td}$ and $V_{ts}$ which are the only CKM
couplings in $\klpnn$. From this point of view it is very natural
to measure the parameter $A$ in a short distance process involving
the top quark and using unitarity of the CKM matrix to find the
value of $\mid V_{cb}\mid$. Moreover this strategy in contrast to
tree-level B-decays is free from hadronic uncertainties. On the
other hand one should keep in mind that this method contains the
uncertainty from the physics beyond the standard model which could
contribute to short distance processes like $\klpnn$. We will return
to this below.

The aim of this letter is to present this strategy for a clean
determination of $\varrho$, $\eta$ and $\mid V_{cb}\mid$ in explicit
terms and to calculate the accuracy for these three parameters which
we think should be achieved in the first decade of the next millennium.
As byproducts we predict very accurate determinations of the top quark
couplings $\mid V_{td} \mid$ and $\mid V_{ts} \mid$ and of the
ratio $\mid V_{ub}/V_{cb}\mid$ which in our opinion are
superior to any other determinations. We derive a set of formulae
which should be useful in future studies of CP asymmetries.
We also present an analysis in which $Br(\kpnn)$ instead of $Br(\klpnn)$
is used. We find that although
the determination of $\mid V_{cb}\mid$ and
$\mid V_{td}\mid$ this way is less accurate, still useful results
for these elements can be obtained.

Our discussion of the Cabibbo-Kobayashi-Maskawa matrix
will be based on the standard parametrization \cite{PDG},
which can equivalently be rewritten in terms of the Wolfenstein
parameters ($\lambda$, $A$, $\varrho$, $\eta$) through the
definitions \cite{BLO}
\b\l{wop}
s_{12}\equiv\lambda \qquad s_{23}\equiv A \lambda^2 \qquad
s_{13} e^{-i\delta}\equiv A \lambda^3 (\varrho-i \eta)      \e
Due to the resulting simplifications, the Wolfenstein
parametrization \cite{WO} is particularly useful when
an expansion in $\lambda=|V_{us}|=0.22$ is performed.
Including next-to-leading terms in $\lambda$ \cite{BLO}
implies that the apex of the reduced unitarity triangle
defined through
\b\l{ut}
\bar\varrho + i \bar\eta\equiv -{V_{ud}V^\ast_{ub}\o
 V_{cd}V^\ast_{cb}}   \e
is with an error of less than 0.1\%
given by
\b\l{1a}
\bar\varrho=\varrho (1-\frac{\lambda^2}{2})
\qquad
\bar\eta=\eta (1-\frac{\lambda^2}{2}) \e
 and not by ($\varrho$, $\eta$) as
usually found in the literature \cite{NQ}. Working in the Wolfenstein
parametrization such a treatment is required if
we aim at a determination of the CKM parameters with a high
precision.

The CP asymmetries in $B^0_{d,s}$-decays to CP eigenstates determine
 $\sin(2\phi_i)$ where $\phi_i=\alpha,~\beta,~\gamma$ are
the angles in the unitarity triangle defined by (\ref{ut}).
Strategies involving
several channels are sometimes necessary in order to remove hadronic
uncertainties
\cite{NQ}.
 $\sin(2\phi_i)$ can be expressed
in terms of $(\bar\varrho,\bar\eta)$ as follows \cite{BLO}
\begin{equation}\label{1}
\sin(2\alpha)=\frac{2\bar\eta(\bar\eta^2+\bar\varrho^2-\bar\varrho)}
  {(\bar\varrho^2+\bar\eta^2)((1-\bar\varrho)^2
  +\bar\eta^2)}
\end{equation}
\begin{equation}\label{2}
\sin(2\beta)=\frac{2\bar\eta(1-\bar\varrho)}{(1-\bar\varrho)^2 + \bar\eta^2}
\end{equation}
\begin{equation}\label{2a}
\sin(2\gamma)=\frac{2\bar\varrho\bar\eta}{\bar\varrho^2+\bar\eta^2}
\end{equation}
Next \cite{BB2}
\begin{equation}\label{3}
Br(\klpnn)=\bar\kappa_L\eta^2\mid V_{cb}\mid^4 X^2(x_t)
\end{equation}
where $x_t=m_t^2/M_W^2$,
\begin{equation}\label{3a}
 \bar\kappa_L
={3\alpha^2 Br(K^+\to\pi^0e^+\nu)\o 2\pi^2\sin^4\Theta_W}
{\tau(K_L)\o\tau(K^+)}=0.354\cdot 10^{-4}
\end{equation}
 and
\b\l{xex0}
X(x) = \eta_X \cdot \frac{x}{8} \left[ - \frac{2+x}{1-x} + \frac{3x
-6}{(1-x)^2} \ln x \right] \qquad \quad \eta_X = 0.985  \e
Here $\eta_X$ is the NLO correction calculated in \cite{BB1,BB2}.
With $m_t\equiv\overline{m_t}(m_t)$ the QCD factor $\eta_X$ is practically
independent of $m_t$.

Using (\ref{1}-\ref{3a}) we can express
 $\bar\varrho$, $\bar\eta$ and
$\mid V_{cb}\mid$ in terms of
\begin{equation}\label{4}
a\equiv \sin(2\alpha)\qquad b\equiv \sin(2\beta)\qquad c\equiv \sin(2\gamma)
\end{equation}
and $Br(\klpnn)$. There are several solutions. We give first only the solution
which is favoured on the basis of what we already know about the CKM matrix.
 Using (\ref{1}-\ref{2a}) we express $\bar\varrho$
 in terms of $\bar\eta$  and $a$, $b$, and $c$ respectively:
\begin{equation}\label{5}
\bar\varrho =
 \frac{1}{2}-\sqrt{\frac{1}{4} -\bar\eta^2+\bar\eta r_{-}(a)}
\quad ,\quad
\bar\varrho = 1-\bar\eta r_{+}(b)\quad ,\quad
\bar\varrho =\bar\eta r_{-}(c)
\end{equation}
where we have introduced
\begin{equation}\label{7}
r_{\pm}(z)=\frac{1}{z}(1\pm\sqrt{1-z^2})
\qquad
z=a,b,c
\end{equation}

For the pairs $(a,b)$, $(b,c)$ and $(a,c)$, assuming $\bar\eta\not=0$,
 we can  then determine
$\bar\eta$ with the result
\begin{equation}\label{8}
\bar\eta=\frac{r_{-}(a)+r_{+}(b)}{1+r_{+}^2(b)}
=\frac{1}{r_{+}(b)+r_{-}(c)}=
\frac{r_{-}(a)+r_{-}(c)}{1+r_{-}^2(c)}
\end{equation}
respectively.
Using (\ref{3}) and (\ref{3a}) we can next determine $\mid V_{cb}\mid$
as follows:
\begin{equation}\label{9}
\mid V_{cb}\mid=\lambda^2\left[ {{\sqrt{B_L}}\o{\eta X(x_t)}}\right]^{1/2}
\qquad
B_L={Br(\klpnn)\o 1.94\cdot 10^{-10}}
\end{equation}
where $\eta$ is to be found from (\ref{1a}) and (\ref{8}).
Note that the factor in front of $\lambda^2$ gives the parameter $A$ in
the Wolfenstein parametrization.
Finally using
\begin{equation}\label{9a}
X(x_t)=0.65\cdot x_t^{0.575}
\end{equation}
which reproduces the function $X(x_t)$ to an accuracy of better than
$0.5\%$ for $150~GeV\leq m_t \leq 190~GeV$ we find a useful formula
\begin{equation}\label{9b}
\mid V_{cb}\mid=39.1\cdot 10^{-3}\sqrt\frac{0.39}{\eta}
\left[ \frac{170 ~GeV}{m_t} \right]^{0.575}
\left[ \frac{Br(\klpnn)}{3\cdot 10^{-11}} \right]^{1/4}
\end{equation}
We note that the weak dependence of $\mid V_{cb}\mid$ on $Br(\klpnn)$
allows to achieve high accuracy for this CKM element even when $Br(\klpnn)$
is known within $5-10\%$ accuracy.

Equations (\ref{5})-(\ref{9}) together with (\ref{1a}) are the main formulae
 which we will use to  determine
$\rho$, $\eta$ and $\mid V_{cb}\mid$. First however we would like to
discuss the remaining solutions.

There are four solutions for $\bar\rho$ coming from (\ref{1}) at fixed
$\sin(2\alpha)$ and $\bar\eta$. They are given by the first formula in
(\ref{5}) with $r_{\pm}(a)$ and with $\pm$ in front of the square root.
The solutions with $+$ in front of the square root can be excluded by
imposing $\mid V_{ub}/V_{cb} \mid \leq 0.10$ in accordance with the
data on semi-leptonic B-decays \cite{STONE}.
 Next there are two solutions for $\bar\rho$
coming from (\ref{2}) given by the second formula in (\ref{5}) with
$r_{\pm}(b)$. The solution with $r_{-}(b)$ violates
$\mid V_{ub}/V_{cb} \mid \leq 0.10$ and consequently there is only one
acceptable solution coming from (\ref{2}). Finally there are two solutions
for $\bar\rho$ from (\ref{2a}) given by the last formula in (\ref{5}) with
$r_{\pm}(c)$.

Retaining the allowed solutions for $\bar\rho$ we find then
the generalization of (\ref{8}) with $(r_{\pm}(a),r_{+}(b))$,
$(r_{+}(b),r_{\pm}(c))$ and $(r_{\pm}(a),r_{\pm}(c))$, in an obvious
notation, respectively. One can then check numerically by varying
$(a,b,c)$ in the full range that
the unique solution with $(r_{-}(a),r_{+}(b))$,
$(r_{+}(b),r_{-}(c))$ and $(r_{-}(a),r_{-}(c))$ as given in (\ref{5})
and (\ref{8}) is obtained if one requires
\begin{equation}\label{19}
\eta>0.20,\qquad -0.20\leq\varrho\leq 0.25,
\qquad 0.06\leq\mid\frac{V_{ub}}{V_{cb}}\mid\leq 0.10
\end{equation}
This range is favoured by a recent analysis of the unitarity triangle
\cite{BLO} and the data on B-decays \cite{STONE}.
We will discuss only this solution in what follows. In the future when
$(a,b,c)$ will be measured one will have to use a similar strategy
to select a unique solution by means of other measurements.
Once this has been done, a precise determination of CKM parameters within
this solution will be possible along the lines discussed here.
We will illustrate this on examples below.

We now turn to a numerical investigation of the formulae above.
In (\ref{3a}) and (\ref{9}) we use
\cite{PDG}
\b\l{las}
\lambda=0.22\quad\alpha=1/128\quad \sin^2\Theta_W=0.23 \e
\b\l{btklp}
Br(K^+\to\pi^0e^+\nu)=4.82\cdot 10^{-2}\qquad
\tau(K_L)/\tau(K^+)=4.18   \e
and assume that the $O(1\%)$ uncertainties in these numerical
values will be reduced in the coming years to the level that they
can be neglected. We will also neglect the small residual scale
ambiguity in $X(x_t)$ \cite{BB2} which can effectively
 be taken into account
by introducing an additional error $\Delta m_t \leq \pm 1~GeV$.

As illustrative examples, let us consider the following three scenarios:

\bigskip
\underline{Scenario I}
\begin{equation}\label{210}
\begin{array}{rclrcl}
\sin(2\alpha) & = &  0.40 \pm 0.08 &
\sin(2\beta)  & = & 0.70 \pm 0.06 \\
Br(\klpnn) & = & (3.0 \pm 0.3)\cdot 10^{-11} &
m_t & = & (170 \pm 5)~GeV \\
\end{array}
\end{equation}

\bigskip
\underline{Scenario II}
\begin{equation}\label{211}
\begin{array}{rclrcl}
\sin(2\alpha) & = &  0.40 \pm 0.04 &
\sin(2\beta)  & = & 0.70 \pm 0.02 \\
Br(\klpnn) & = & (3.00 \pm 0.15)\cdot 10^{-11} &
m_t & = & (170 \pm 3)~GeV \\
\end{array}
\end{equation}

\bigskip
\underline{Scenario III}
\begin{equation}\label{212}
\begin{array}{rclrcl}
\sin(2\alpha) & = &  0.40 \pm 0.02 &
\sin(2\beta)  & = & 0.70 \pm 0.01 \\
Br(\klpnn) & = & (3.00 \pm 0.15)\cdot 10^{-11} &
m_t & = & (170 \pm 3)~GeV \\
\end{array}
\end{equation}
The accuracy in the scenario I should be achieved at B-factories
\cite{BAB}, HERA-B \cite{AL},
at KAMI \cite{AR} and at KEK \cite{ISS}.
 Scenarios II and
III correspond to B-physics at Fermilab during the Main Injector
era \cite{FNAL} and at LHC
\cite{CA,B93} respectively.
At that time an improved measurement of $Br(\klpnn)$ should be aimed for.
The values of $m_t$ assumed here are in the ball park of the most recent
results of the CDF collaboration \cite{CDF}. Since in accordance with
the QCD corrections in \cite{BB2} we use here the current top quark mass
at the scale  $m_t$, our values correspond to $m_t^{phys}=(177\pm 3)~GeV$.
This should be compared with $m_t^{phys}=(174\pm 16)~GeV$ reported by CDF
\cite{CDF}.

The results that would be obtained in these scenarios for $\rho$, $\eta$,
$\mid V_{cb}\mid$,\\ $\mid V_{ub}/V_{cb}\mid$, $\mid V_{td}\mid$,
$\mid V_{ts}\mid$ and $\sin(2\gamma)$ are collected in table 1.
 To this end we have used \cite{BLO}
\begin{equation}\label{11}
\mid V_{td}\mid=\mid V_{cb}\mid \lambda \sqrt{(1-\bar\varrho)^2 +\bar\eta^2}
\qquad
\left| \frac{V_{ub}}{V_{cb}} \right|=\lambda \sqrt{\varrho^2 +\eta^2}
\end{equation}
and the standard expression for $\mid V_{ts}\mid$ \cite{PDG}.

\begin{table}
\begin{center}
\begin{tabular}{|c||c||c|c|c|}\hline
& Central &$I$&$II$&$III$\\ \hline
$\rho$ &$0.072$ &$\pm 0.040$&$\pm 0.016$ &$\pm 0.008$\\ \hline
$\eta$ &$0.389$ &$\pm 0.044$ &$\pm 0.016$&$\pm 0.008$ \\ \hline
$\mid V_{cb}\mid/10^{-3}$ &$39.2$ &$\pm 3.9$ &$\pm 1.7$&$\pm 1.3$\\ \hline
$\mid V_{ub}/V_{cb}\mid$ &$0.087$ &$\pm 0.010$ &$\pm 0.003$&$\pm 0.002$
 \\ \hline
$\mid V_{td}\mid/10^{-3}$ &$8.7$ &$\pm 0.9$ &$\pm 0.4$ &$\pm 0.3$ \\
 \hline
$\mid V_{ts}\mid/10^{-3}$ &$38.4$ &$\pm 3.8$ &$\pm 1.7$&$\pm 1.3$ \\ \hline
$\sin(2\gamma)$ &$0.35$ &$\pm 0.15$ &$\pm 0.07$&$\pm 0.04 $ \\ \hline
$B_K$ &$0.83$ &$\pm 0.17$ &$\pm 0.07$&$\pm 0.06$ \\ \hline
$F_B\sqrt{B_B}$ &$200$ &$\pm 19$ &$\pm 8$&$\pm 6$ \\ \hline
\end{tabular}
\end{center}
\centerline{}
{\bf Table 1:} Determinations of various parameters in scenarios I-III with
$F_B\sqrt{B_B}$ given in $MeV$. The errors have been symmetrized so that
sometimes the central values do not exactly correspond to the central
values of
the input parameters.
\end{table}

In table 2 we show the results for the corresponding scenarios I'-III'
in which
$\sin(2\gamma)$ is used instead of $\sin(2\alpha)$. Here we set
\begin{equation}\label{222}
\sin(2 \beta) = \left\{ \begin{array}{rc}
0.60 \pm 0.06 & (\rm{I'})  \\
0.60 \pm 0.02  & (\rm{II'}) \\
0.60 \pm 0.01 & (\rm{III'})
\end{array}\right.
\qquad
\sin(2 \gamma) = \left\{ \begin{array}{rc}
0 \pm 0.08 & (\rm{I'})  \\
0 \pm 0.04  & (\rm{II'}) \\
0 \pm 0.02 & (\rm{III'})
\end{array}\right.
\end{equation}
Furthermore we use $m_t = 180~GeV$ and $Br(\klpnn)=3.2\cdot 10^{-11}$
with the errors as in scenarios I-III. We are aware of the fact that
 the accuracy for $\sin(2\gamma)$ assumed in scenarios II' and III' will
 be difficult to achieve \cite{NQ,GRWY}
but we show the results of this exercise to
 further motivate the efforts in this field. For $\rho$ and $\eta$ we
give three decimal places in order to show better the differences between
various cases.
\begin{table}
\begin{center}
\begin{tabular}{|c||c||c|c|c|}\hline
& Central &$I'$&$II'$&$III'$\\ \hline
$\rho$ &$0.000$ &$\pm 0.016$&$\pm 0.007$ &$\pm 0.004$\\ \hline
$\eta$ &$0.342$ &$\pm 0.048$ &$\pm 0.017$&$\pm 0.009$ \\ \hline
$\mid V_{cb}\mid/10^{-3}$ &$41.2$ &$\pm 4.5$ &$\pm 1.9$&$\pm 1.4$\\ \hline
$\mid V_{ub}/V_{cb}\mid$ &$0.075$ &$\pm 0.010$ &$\pm 0.004$&$\pm 0.002$
 \\ \hline
$\mid V_{td}\mid/10^{-3}$ &$9.6$ &$\pm 0.9$ &$\pm 0.4$ &$\pm 0.3$ \\
 \hline
$\mid V_{ts}\mid/10^{-3}$ &$40.2$ &$\pm 4.4$ &$\pm 1.8$&$\pm 1.4$ \\ \hline
$\sin(2\alpha)$ &$0.60$ &$\pm 0.12$ &$\pm 0.05$&$\pm 0.03 $ \\ \hline
$B_K$ &$0.69$ &$\pm 0.15$ &$\pm 0.06$&$\pm 0.05$ \\ \hline
$F_B\sqrt{B_B}$ &$174$ &$\pm 15$ &$\pm 6$&$\pm 5$ \\ \hline
\end{tabular}
\end{center}
\centerline{}
{\bf Table 2:} Determinations of various parameters in scenarios I'-III'.
\end{table}
One can easily check that other possible solutions for $(\varrho,\eta)$
can be excluded. Choosing for instance the solution $(r_+(a),r_+(b))$ in
scenarios I-III one finds $(\varrho,\eta)\approx (-1.6,1.1)$ and
$\mid V_{ub}/V_{cb}\mid \approx 0.43$ which is much too large. Similarly
the solution $(r_+(b),r_+(c))$ in
scenarios I'-III' gives $(\varrho,\eta)\approx (1.0,0)$ and
$\mid V_{ub}/V_{cb}\mid \approx 0.22$ which is also too large. In the
same manner all solutions discussed above except for the one presented
in the tables can be excluded.

Tables 1 and 2 show very clearly the potential of CP asymmetries
in B-decays and of $\klpnn$ in the determination of CKM parameters.
It should be stressed that this high accuracy is not only achieved
because of our assumptions about future experimental errors in the
scenarios considered but also because $\sin(2\alpha)$ and $\sin(2\gamma)$
are very sensitive functions of $\varrho$ and $\eta$ \cite{BLO}
and $Br(\klpnn)$
depends strongly on $\mid V_{cb}\mid$. In particular as seen in table 2
the determination of $\sin(2\gamma)$ may offer a very precise
 measurement of $\varrho$.

 The accuracy in scenarios I and I'
corresponds roughly to the cases considered in \cite{BLO}, where however
$\klpnn$ has not been discussed. There also the prospects of the
determination of the unitarity
triangle using $B^{\circ}_d-\bar B^{\circ}_d$ mixing and the parameter
$\varepsilon_K$ have been analyzed. The results of \cite{BLO} show that in
such an analysis it
will be very difficult to determine $\varrho$ and $\eta$ to better than
$\Delta \varrho=\pm 0.10 $ and $\Delta\eta=\pm 0.05$ which should be
contrasted with the accuracy expected here. Similarly because of
theoretical uncertainties it is at present difficult to imagine that in
the tree level B-decays a better accuracy than
$\Delta \mid V_{cb}\mid =\pm 2\cdot 10^{-3}$ and
$\Delta \mid V_{ub}/V_{cb}\mid=\pm 0.01$ could be achieved.

In tables 1 and 2 we have also shown the values of the non-perturbative
parameters $B_K$ and  $F_B \sqrt{B_B}$ which can be
extracted from the data on $\varepsilon_K$ and
$B^{\circ}_d-\bar B^{\circ}_d$ mixing once the CKM parameters have been
determined in the scenarios considered. To this end
$x_d=0.72$ and $\tau(B)=1.5~ps$ have been assumed. The errors
on these two quantities should be negligible at the end of this
millennium. Note that the resulting central values for $B_K$ in tables
 1 and 2
are close to the lattice \cite{SHARP} and $1/N$ \cite{BBG}
results respectively..

We have checked that similar patterns of uncertainties
emerge for different central input parameters.
Needless to say when the scenarios presented here will become a reality
one will have to make sure that the uncertainties present in the input
parameters in (\ref{las}) and (\ref{btklp})
have been reduced to the desired level.

It is instructive to investigate whether the use of another
short distance decay
$\kpnn$ instead of $\klpnn$ would also give interesting results for
$V_{cb}$ and $V_{td}$. $\kpnn$ is CP conserving and receives also
contributions from internal charm exchanges. This introduces the
dependence on $\Lambda_{\overline{MS}}$ and $m_c$ but otherwise
this decay is known to be theoretically very clean.
In particular the long distance contributions to
$K^+ \rightarrow \pi^+ \nu \bar{\nu}$ have been
considered in \cite{RS,HL,LW} and found to be very small: two to three
orders of magnitude smaller than the short distance contribution
at the level of the branching ratio.
Moreover
in contrast to $\klpnn$, the decay $\kpnn$ could be
observed already in the coming years at AGS in Brookhaven.
 Using the expressions in \cite{BB4} and \cite{BLO}
we find instead of (\ref{9})
\begin{equation}\label{27}
\mid V_{cb}\mid=
\lambda^2\left[ {{\sqrt{B_+(\eta^2+b^2_1)-b^2_0\eta^2}-b_0 b_1}
\o{(\eta^2+b^2_1) X(x_t)}}\right]^{1/2}
\end{equation}
where
\begin{equation}\label{28}
B_+={Br(\kpnn)\o 4.64\cdot 10^{-11}}
\qquad
b_0=(1-\frac{\lambda^2}{2})P_0(K^+)
\end{equation}
and
\begin{equation}\label{28a}
b_1=1-\varrho+\lambda^2 (2\varrho-\varrho^2-\eta^2-\frac{1}{2})+
O(\lambda^4)
\end{equation}
Here $P_0(K^+)$ represents the charm contribution to $\kpnn$
calculated including next-to-leading QCD corrections in \cite{BB3}.
For $200~MeV\leq \Lambda_{\overline{MS}}\leq 350~MeV$
 $1.25~GeV\leq m_c\leq 1.35~GeV$,
 one has $P_0(K^+)=0.40\pm 0.09$ \cite{BB3,BB4} where also
the residual uncertainty due to the choice of the renormalization scale
 $\mu$ has been taken into account. Here $m_c$ stands for the running
charm quark mass at the $m_c$ scale.

 We again consider scenarios I-III with
$Br(\kpnn)= (1.0\pm 0.1)\cdot 10^{-10}$ for the scenario I and
$Br(\kpnn)= (1.0\pm 0.05)\cdot 10^{-10}$ for scenarios II and III
in place of $Br(\klpnn)$ with all other input parameters unchanged.
The results for $\varrho$, $\eta$, $\mid V_{ub}/V_{cb}\mid$ and
$\sin(2\gamma)$ remain of course unchanged. In table 3 we show the
results for $V_{cb}$, $V_{td}$ and $F_B\sqrt{B_B}$ . We observe that
due to the uncertainties present in the charm contribution to
$\kpnn$, which was absent in $\klpnn$, the determinations of
 $\mid V_{cb}\mid$,
 $\mid V_{td}\mid$ and $F_B\sqrt{B_B}$ are less accurate. On the
other hand,
we note in the case of $\mid V_{td} \mid $ that due to an
 approximate cancellation of the square root in (\ref{11}) by
the denominator in (\ref{27}), the uncertainties due to
the errors in $\eta$ and $\varrho$ are considerably reduced.
Consequently when compared with $\klpnn$ the reduction of the
accuracy in $V_{td}$ is less pronounced than in the case of $V_{cb}$.
 If the uncertainties due to the charm mass
and $\Lambda_{\overline{MS}}$ are removed one day, only the uncertainty
related to $\mu$ will remain in $P_0(K^+)$ giving
$\Delta P_0(K^+)=\pm 0.03$ \cite{BB3}.
 In this case the results in parentheses
in table 3 would be found.

\begin{table}
\begin{center}
\begin{tabular}{|c||c||c|c|c|}\hline
& Central &$I$&$II$&$III$\\ \hline
$\mid V_{cb}\mid/10^{-3}$ &$41.2$ &$\pm 4.3~(3.2)$ &$\pm 3.0~(1.9)$&
$\pm 2.8~(1.8)$\\ \hline
$\mid V_{td}\mid/10^{-3}$ &$9.1$ &$\pm 0.9~(0.7)$ &$\pm 0.6~(0.4)$&
$\pm 0.6~(0.4)$ \\
 \hline
$F_B\sqrt{B_B}$ &$190$ &$\pm 17~(12)$ &$\pm 12~(7)$&$\pm 12~(7)$ \\ \hline
\end{tabular}
\end{center}
\centerline{}
{\bf Table 3:} Determinations of various parameters in scenarios I-III
using $\kpnn$ instead of $\klpnn$. The values in parentheses show
the situation when the uncertainties in $m_c$ and $\Lambda_{\overline{MS}}$
are not included.
\end{table}

Let us finally summarize the main aspects of this letter.
\\
We have pointed out that the measurements of the CP asymmetries in neutral
B-decays together with a measurement of $Br(K_L\to \pi^\circ\nu\bar\nu)$
and the known value of $\mid V_{us}\mid$ offer a precise determination
of {\it all} elements
of the Cabibbo-Kobayashi-Maskawa matrix without essentially
any hadronic uncertainties.
$\klpnn$ proceeds almost entirely through direct CP violation and is known
to be a very useful decay for the determination of $\eta$. However due
to the strong dependence on $V_{cb}$ this determination cannot fully
compete with the one which can be achieved using CP asymmetries in
B-decays. As a recent analysis \cite{BB4} shows it will be difficult
to reach $\Delta \eta=\pm 0.03$ this way if $\mid V_{cb}\mid$ is determined
in tree level B-decays. Our strategy then is to find $\eta$ from
CP asymmetries in B decays and use $\klpnn$ to determine $\mid V_{cb}\mid$.
To our knowledge no other decay can determine $\mid V_{cb}\mid$ as
cleanly as this one.

We believe that the strategy presented here is the theoretically cleanest
way to establish the precise values of the CKM parameters. The ratio
$x_d/x_s$ of $B^0_d-\bar B^0_d$ to $B^0_s-\bar B^0_s$ mixings,
$Br(\kpnn)$ and the parity violating asymmetry ($\Delta_{LR}$)
in $K^+\to\pi^+\mu\bar\mu$ \cite{SW1,SW2,GT1}
are also theoretically rather clean and are useful in this respect.
In particular as demonstrated in \cite{BB4}, $\kpnn$ together with
$\klpnn$ offers a respectable determination of $\sin(2\beta)$.
However from the theoretical point
of view the determinations of this type cannot compete with the strategy
 considered here.
$x_d/x_s$ is subject to uncertainties related to SU(3) flavour breaking
effects which will probably be difficult to bring below $5\%$.
Moreover the dependence on $A$ cancells in this ratio and consequently
$V_{cb}$ cannot be determined.
$Br(\kpnn)$ is more sensitive to $\Lambda_{\overline{MS}}$ than
$Br(\klpnn)$ and in addition receives a few $\%$ uncertainty from the
error in the charm quark mass \cite{BB3}. As we have seen above these
uncertainties lower the precision of the determination of $V_{cb}$ and
$V_{td}$
compared to $\klpnn$, although as shown in table 3 interesting
results for these elements  can still be obtained.
Similar comments can be made about
$\Delta_{LR}$ for which in addition possible
long distance
contributions at a few $\%$ level cannot be excluded \cite{SW2}.

On the other hand once $\rho$, $\eta$ and $\mid V_{cb}\mid$ (or A) have
been precisely determined as discussed here, it is clear that $x_d/x_s$,
$Br(\kpnn)$ and $\Delta_{LR}$ can be rather accurately predicted and
confronted with future experimental data. Such confrontations would
offer excellent tests of the standard model and could possibly give
signs of a new physics beyond it.

Of particular interest will also be the comparision of $\mid V_{cb}\mid$
determined as suggested here with the value of this CKM element extracted
from tree level semi-leptonic  B-decays. Since in contrast to
$\klpnn$, the tree-level decays are to an excellent approximation
insensitive to any new physics contributions from very high energy scales,
the comparison of these two determinations of $\mid V_{cb}\mid$ would
be an excellent test of the standard model and of a possible physics
beyond it. Also the values of $\mid V_{ub}/V_{cb}\mid$ from tree-level
B-decays, which are subject to hadronic uncertainies  larger
than in the case of $V_{cb}$, when compared with the clean determinations
suggested here could teach us about the realiability of non-perturbative
methods. The same applies to the quantities like $x_d$ and the CP violating
parameter $\varepsilon_K$ which are subject to uncertainties present in
the non-perturbative parameters $F_B \sqrt{B_B}$ and $B_K$ respectively.
In tables 1-3 we have shown the values of these parameters which can be
extracted from the data in the scenarios considered.

It is also clear that once the accuracy for CKM parameters presented
here has been attained, also detailed tests of proposed schemes
for quark matrices \cite{HALL,ROSS} will be possible.

Precise determinations of all CKM parameters without hadronic uncertainties
 along the lines suggested
here can only be realized if the measurements of CP asymmetries in
B-decays and the measurements of $Br(\klpnn)$ and $Br(\kpnn)$
can reach the desired accuracy.
All efforts should be made to achieve this goal.
\vskip0.5cm

I would like to thank Gerhard Buchalla, Robert Fleischer and Markus
Lautenbacher for critical reading of the manuscript and Markus
Lautenbacher for checking my numerical analysis. I also thank
Hubert Kroha and Karl Jakobs for illuminating discussions about
future B physics experiments.

\vfill\eject


\vfill
\newpage

\begin{thebibliography}{99}
\bibitem{CAB}
N. Cabibbo, {\sl Phys. Rev. Lett.} {\bf 10} (1963) 531.
\bibitem{KM}
M. Kobayashi and K. Maskawa, {\sl Prog. Theor. Phys.} {\bf 49} (1973) 652.
\bibitem{LR}
H. Leutwyler and M. Roos, {\sl Zeitschr. f.  Physik} {\bf C25} (1984) 91.
\bibitem{DHK}
J.F. Donoghue, B.R. Holstein and S.W. Klimt, {\sl Phys. Rev.} {\bf D 35}
 (1987) 934.
\bibitem{MAR}
W.J. Marciano, {\sl Annu. Rev. Nucl. Part. Sci.} {\bf 41} (1991) 469 and
talk given at the Symposium on Heavy Flavours, Montreal, Canada, 1993.
\bibitem{NQ}
Y. Nir and H.R. Quinn in " B Decays ", ed S. Stone (World Scientific, 1992),
p. 362; I. Dunietz, ibid p.393 and refs. therein.
\bibitem{AKL}
R. Aleksan, B. Kayser and D. London, NSF-PT-93-4 NSF-PT-94-2
and refs. therein.
\bibitem{WO}
L. Wolfenstein, {\sl Phys. Rev. Lett.} {\bf 51} (1983) 1945.
\bibitem{BLO}
A.J. Buras, M.E. Lautenbacher and G. Ostermaier,
MPI-Ph/94-14, TUM-T31-57/94.
\bibitem{LI}
L.S. Littenberg, {\sl Phys. Rev.} {\bf D 39} (1989) 3322.
\bibitem{BB1}
G. Buchalla and A.J. Buras,
{\sl Nucl. Phys.} {\bf B 398} (1993) 285.
\bibitem{BB2}
G. Buchalla and A.J. Buras,
{\sl Nucl. Phys.} {\bf B 400} (1993) 225.
\bibitem{DDG}
C.O. Dib, I. Dunietz and F.J. Gilman, {\sl Mod. Phys. Lett.} {\bf A 6}
(1991) 3573.
\bibitem{BAB}
BaBar Collaboration -- Status report, SLAC-419 (June 1993), \\
BELLE Collaboration (KEKB) -KEK-94-2 (April 1994).
\bibitem{AL}
H. Albrecht et al. (HERA-B), DESY-PRC 92/04 (1992).
\bibitem{FNAL}
CDF Collaboration and Others, CDF/DOC/ADVISORY/PUBLIC/2436 EOI (May 1994);
R. Edelstein et al. EOI (May 1994).
\bibitem{CA}
S. Erhan et al. (COBEX), {\sl Nucl. Instr. and Methods} {\bf A333}
(1993) 101;\\ ATLAS Collaboration, CERN/LHCC/93-53;
\bibitem{B93}
Workshop on Beauty 93, Liblice Castle, Czech Republic, 1993,
ed. P. Schlein, {\sl Nucl. Instr. and Methods} {\bf A333}
(1993) 1-230.
\bibitem{AR}
K. Arisaka et al., KAMI conceptual design report, FNAL, June 1991
\bibitem{ISS}
T. Inagaki, T. Sato and T. Shinkawa, Experiment to search for the
decay $\klpnn$ at KEK $12 GeV$ Proton Synchrotron, 30 Nov. 1991.
\bibitem{PDG}
Particle Data Group, {\sl Phys. Rev.} {\bf D 45} (1992), No.11 part II.
\bibitem{STONE}
S. Stone, in B Decays II ed. by S. Stone, World Scientific; HEPSY 93-11.
\bibitem{CDF}
F. Abe et al. (CDF Collaboration) FERMI-PUB-94/097-E.
\bibitem{GRWY}
M. Gronau and D. Wyler,{\sl Phys. Lett.} {\bf B 265} (1991) 172.
\bibitem{SHARP}
S.R. Sharpe, UW/PT-93-25 and references therein.
\bibitem{BBG}
W. A. Bardeen, A.J. Buras, and J.-M. Gerard,
 {\sl Phys. Lett.} {\bf B 211} (1988) 343.
\bibitem{RS}
D. Rein and L.M. Sehgal, {\sl Phys. Rev.} {\bf D 39} (1989) 3325.
\bibitem{HL}
J.S. Hagelin and L.S. Littenberg, {\sl Prog. Part. Nucl. Phys.}
{\bf 23} (1989) 1.
\bibitem{LW}
M. Lu and M.B. Wise, CALT-68-1911 (1994).
\bibitem{BB4}
G. Buchalla and A.J. Buras, MPI-PhT/94-19, TUM-T31-59/94.
\bibitem{BB3}
G. Buchalla and A.J. Buras,
{\sl Nucl. Phys.} {\bf B 412} (1994) 106.
\bibitem{SW1}
M. Savage and M. Wise, {\sl Phys. Lett.} {\bf B 250} (1990) 151.
\bibitem{SW2}
M. Lu, M. Wise and M. Savage,
{\sl Phys. Rev.} {\bf D 46} (1992) 5026.
\bibitem{GT1}
G. Belanger, C.Q. Geng and P. Turcotte,
 {\sl Nucl. Phys.} {\bf B 390} (1993) 253.
\bibitem{HALL}
S. Dimopoulos, L.J. Hall and S. Raby,{\sl Phys. Rev.}
 {\bf D 45} (1992) 4192.
\bibitem{ROSS}
P. Ramond, R.G Roberts and G.G. Ross, {\sl Nucl. Phys.} {\bf B 406}
 (1993) 19.
\end{thebibliography}
\end{document}